\documentclass[usenatbib]{mn2e}

\usepackage{amsmath}
\usepackage{graphicx}
\usepackage{float}
\usepackage{wrapfig}
\usepackage{natbib}

\begin{document}

\title{ The Helium Abundance in the Ejecta of U Scorpii}

\author[M. P. Maxwell et al.]{M. P. Maxwell,$^1$\thanks{Email: mpmaxwell@uclan.ac.uk} M. T. Rushton,$^1$ M. J. Darnley,$^2$ H. L. Worters$^3$, M. F. Bode,$^2$
\newauthor  A. Evans,$^4$ S. P. S. Eyres,$^1$ M. B. N. Kouwenhoven,$^5$ F. M. Walter,$^6$ and B. J. M. Hassall$^1$\\
$^1$Jeremiah Horrocks Institute, University of Central Lancashire, Preston, PR1 2HE \\
$^2$Astrophysics Research Institute, Liverpool John Moores University, Birkenhead, CH41 1LD \\
$^3$South African Astronomical Observatory, PO Box 9, 7935 Observatory, South Africa \\
$^4$Astrophysics Group, Keele University, Keele, Staffordshire, ST5 5BG \\
$^5$Kavli Institute for Astronomy and Astrophysics, Peking University, Yi He Yuan Lu 5, Haidian Qu, Beijing 100871, China \\
$^6$Department of Physics and Astronomy, Stony Brook University, Stony Brook, NY 11794-3800, USA}

\maketitle
\begin{abstract}
U Scorpii is a recurrent nova which has been observed in outburst on 10 occasions, most recently in 2010. We present near-infrared and optical spectroscopy of the 2010 outburst of U Sco. The reddening of U Sco is found to be $E(B-V) = 0.14\pm0.12$, consistent with previous determinations, from simultaneous optical and near-IR observations. The spectra show the evolution of the line widths and profiles to be consistent with previous outbursts. Velocities are found to be up to 14000\,kms$^{-1}$ in broad components and up to 1800\,kms$^{-1}$ in narrow line components, which become visible around day 8 due to changes in the optical depth. From the spectra we derive a helium abundance of $N$(He)/$N$(H)$ = 0.073\pm0.031$ from the most reliable lines available; this is lower than most other estimates and indicates that the secondary is not helium-rich, as previous studies have suggested. 
\end{abstract}


\section{Introduction}
U Scorpii is a recurrent nova (RN) which has undergone recorded outbursts in 1863, 1906, 1917, 1936, 1945, 1969, 1979, 1987, and 1999 \citep{schaeferlong} before a further outburst peaked on 2010 January $28.19\pm0.17$ UT \citep{schaefer}. The mean recurrence time is 10.3 years \citep{schaeferlong}. U Sco is an eclipsing binary with an orbital inclination of $\sim82^{\circ}$ \citep{thoroughgood} and is semi-detached with an orbital period of $\simeq$1.23 days \citep{schaefer&ringwald}. U Sco consists of a white dwarf (WD) primary and a probable subgiant secondary with a spectral type in the range K2 \citep{anupama} to G0 \citep{hanes}, with the white dwarf having a mass close to the Chandrasekhar limit ($\simeq 1.37 M_{\odot}$) \citep{thoroughgood}. The system is at a distance of $12\pm2$\,kpc and is far out of the galactic plane at a height of $\sim4.5$\,kpc \citep{schaeferlong}.

\cite{starrfield} interprets the outbursts as being due to a thermonuclear runaway (TNR) on the surface of the white dwarf. The TNR occurs when the temperature and density at the base of the layer accreted from the secondary reach critical values of  $\simeq 10^{8}$\,K and $\simeq 10^{19}$\,Nm$^{-2}$ respectively \citep{starrfield}. The recurrence time scale is consistent with the nova outburst models of \cite{yaron}. The energy released from the hydrogen burning is sufficient to allow heavier elements to undergo nuclear fusion. This continues until the energy generation is limited by the long, temperature independent half-lives of some isotopes involved in the CNO cycle such as $^{14}$O and $^{13}$N.

The 2010 outburst was anticipated by \cite{schaefer04} who planned a multiwavelength observing programme ahead of time; this resulted in the 2010 outburst of U Sco having the best temporal coverage of any nova event so far. This led to the discovery of new phenomena such as aperiodic dips in the light curve and flares (\cite{pagnotta}; Schaefer et al. in prep). The apparent onset of optical flickering on day 8 of the outburst \citep{worters} has been identified as an example of such a flare. The spectral evolution between outbursts is consistent, as is the photometric evolution. \cite{banerjee} present near-IR spectra of the outburst which show broad H\,{\sc i}, He\,{\sc i}, He\,{\sc ii}, and O\,{\sc i} emission lines. The H\,{\sc i} lines give an upper limit on the ejected mass of $9.71\pm9.29 \times 10^{-5}M_{\odot}$ \citep{banerjee}. The helium abundance, $N$(He)/$N$(H), of U Sco is highly uncertain \citep{diaz} with estimates ranging from 0.16 \citep{iijima} to 4.5 \citep{evans}. An accurate determination of the helium abundance in U Sco is necessary as some studies have suggested that, unlike classical novae, the WD is accreting helium-rich material from the secondary star. As a result, the physics of the TNR has been modified by \cite{Starrfield2} for the case of U Sco. To explain the presence in the system of a helium-rich donor star, \cite{hachisukato} proposed an evolutionary sequence in which a helium-rich primary filled its Roche lobe and transfered material to the secondary in the post common envelope phase. However, \cite{truran} showed that a bright nova outburst is only possible on a 1.38 solar mass WD if H/He $\le 1$. This agrees with evolutionary calculations by \cite{sarna}, who concluded that any helium enrichment is from the WD.

Here we present near-infrared (IR) and optical spectroscopy of the latest outburst obtained at the New Technology Telescope (NTT), Liverpool Telescope (LT, \cite{steele}), Cerro Tololo Inter-American Observatory (CTIO), and South African Astronomical Observatory (SAAO) from which we derive the reddening, helium abundance, and broad and narrow component line widths.

\section{Observations}
We take the observed maximum at 2010 January 28.19 to be day 0 of the outburst and all epochs are relative to this time. Near-IR spectroscopy of U Sco was obtained on days 5.41 and 9.43 at the NTT using the Son Of ISAAC (SOFI) spectrograph. The spectral resolution is $\sim1000$. For more details see \cite{banerjee}. An optical spectrum was also obtained on day 9.43 using the Cassegrain spectrograph on the 1.5 metre SMARTS Telescope at CTIO at a resolution of $\sim$1500. Optical spectra were obtained from 1.93 to 11.93 days after outburst using the Cassegrain spectrograph on the SAAO 1.9 metre telescope with spectral resolution $\sim$1000. Optical spectra were also obtained from 6.81 to 12.81 days after outburst using the FRODOSpec \citep{morales} spectrograph at LT at a resolution of $\sim5400$.

\section{Results}
Figures 1 and 2 show the dereddened optical spectra of U Sco covering most of the first 13 days of the outburst. The strong emission lines are due to H\,{\sc i}, He\,{\sc i}, N\,{\sc iii}, and, at later times, He\,{\sc ii}. The He\,{\sc i} lines fade as the He\,{\sc ii} lines develop and are no longer detectable by day 11.81. N\,{\sc iii} and H$\gamma$ also fade and are very weak or undetectable by this time. H$\delta$ and He\,{\sc ii} 4686{\AA} are both blended with N\,{\sc iii}. Figures 3 and 4 show near-IR spectra of U Sco; they show the Paschen series of hydrogen in emission along with O\,{\sc i}, He\,{\sc i}, and He\,{\sc ii} emission lines. Paschen $\gamma$ and Paschen $\delta$ are blended with He\,{\sc i} and He\,{\sc ii} respectively. Figure 3 also shows a spectrum from the outburst in 1999 taken at a similar time after maximum to the first of our NTT spectra; the spectra are very similar with the same emission features present and similar relative line strengths.

\begin{center}
\begin{table}
\caption{Observing Log}

\begin{tabular}{lcc}\hline
Day & Wavelength range ({\AA}) & Facility\\ \hline
1.93 & 3500-7250 & SAAO\\
4.93 & 3500-7250 & SAAO\\
5.41 & 9950-24000 & NTT\\
5.93 & 3500-7250 & SAAO\\
6.81 & 3900-5200, 5700-7900 & LT\\
7.81 & 3900-5200, 5700-7900 & LT\\
7.93 & 3500-7250 & SAAO\\
8.81 & 3900-5200, 5700-7900 & LT\\
8.93 & 3500-7250 & SAAO\\
9.43 & 4000-4800, 9950-24000 & CTIO, NTT\\
9.93 & 3500-7250 & SAAO\\
10.93 & 3500-7250 & SAAO\\
11.81 & 3900-5200, 5700-7900 & LT\\
11.93 & 3500-7250 & SAAO\\
12.81 & 3900-5200, 5700-7900 & LT\\ \hline

\end{tabular}
\end{table}
\end{center}

\begin{center}
\begin{figure*}
\includegraphics{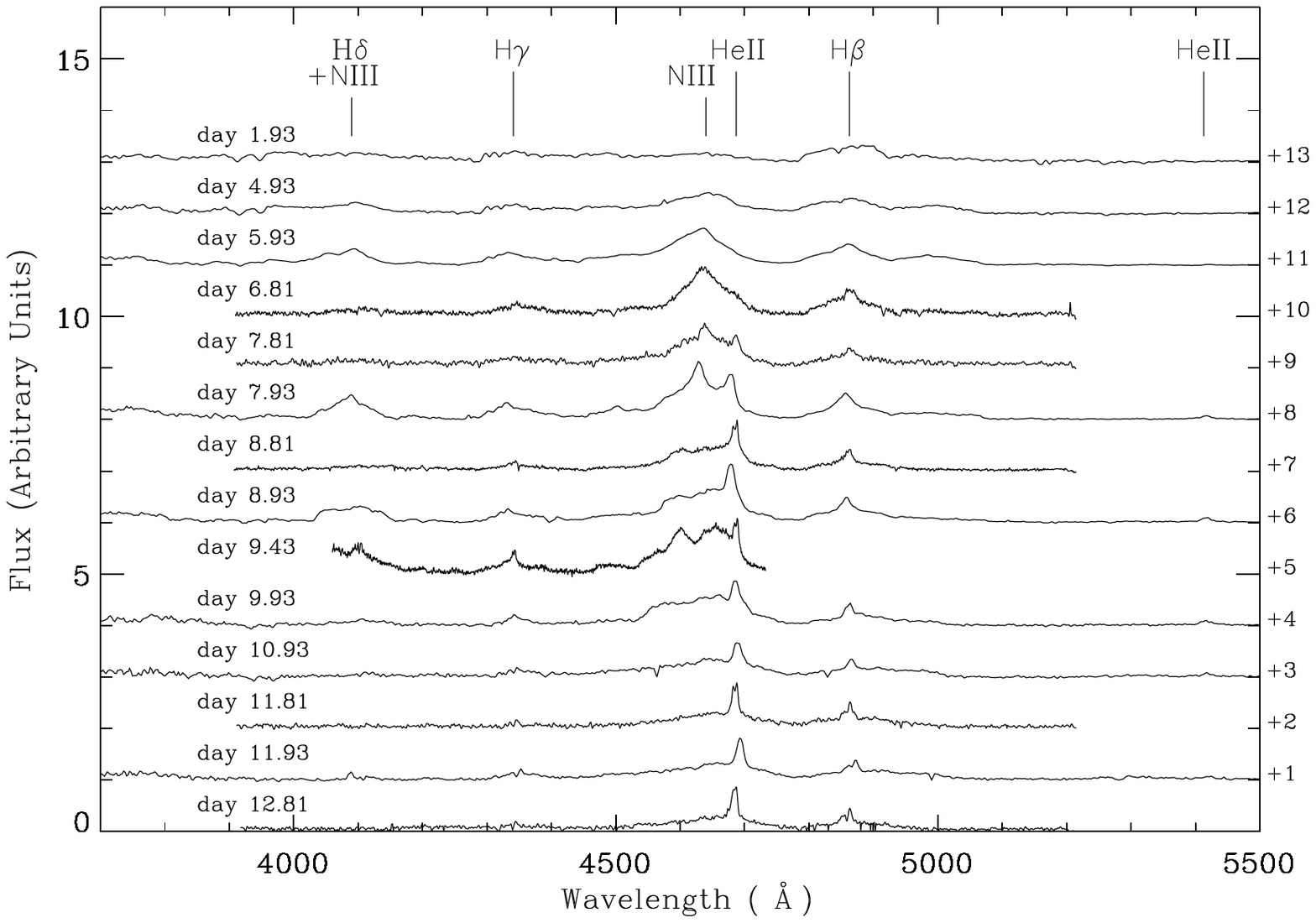}
\caption{Spectra of U Sco taken from day 1.93 to 12.81 at LT (days ending .81), CTIO (day 9.43), and SAAO (days ending .93). The spectra are offset as indicated and the flux is in arbitrary units. LT spectra are smoothed with a Gaussian profile.}
\end{figure*}
\end{center}

\begin{center}
\begin{figure*}
\includegraphics{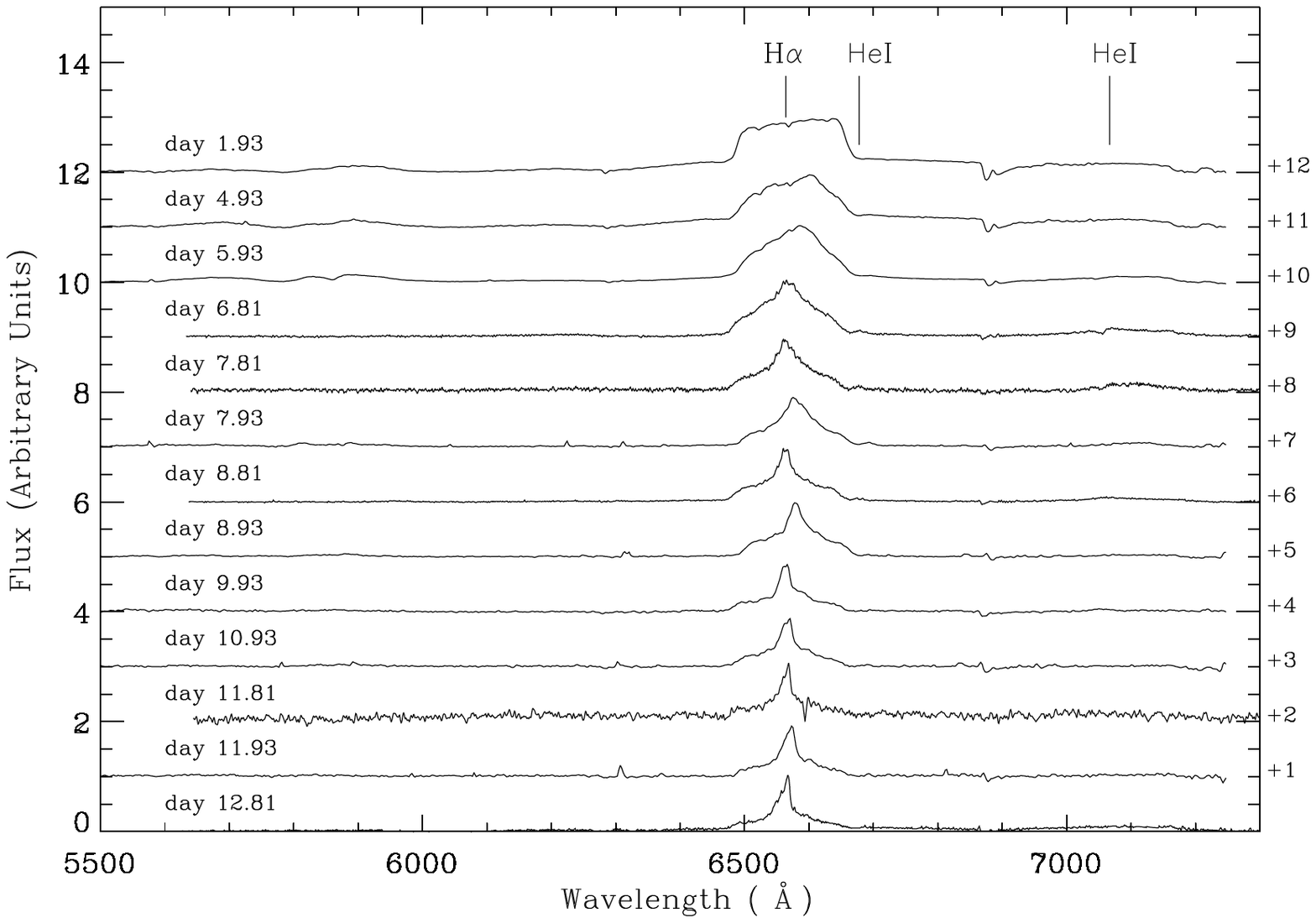}
\caption{Spectra of U Sco taken from day 1.93 to 12.81 at LT (days ending .81) and SAAO (days ending .93). The spectra are offset as indicated and the flux is in arbitrary units. LT spectra are smoothed with a Gaussian profile.}
\end{figure*}
\end{center}

\begin{center}
\begin{figure*}
\includegraphics{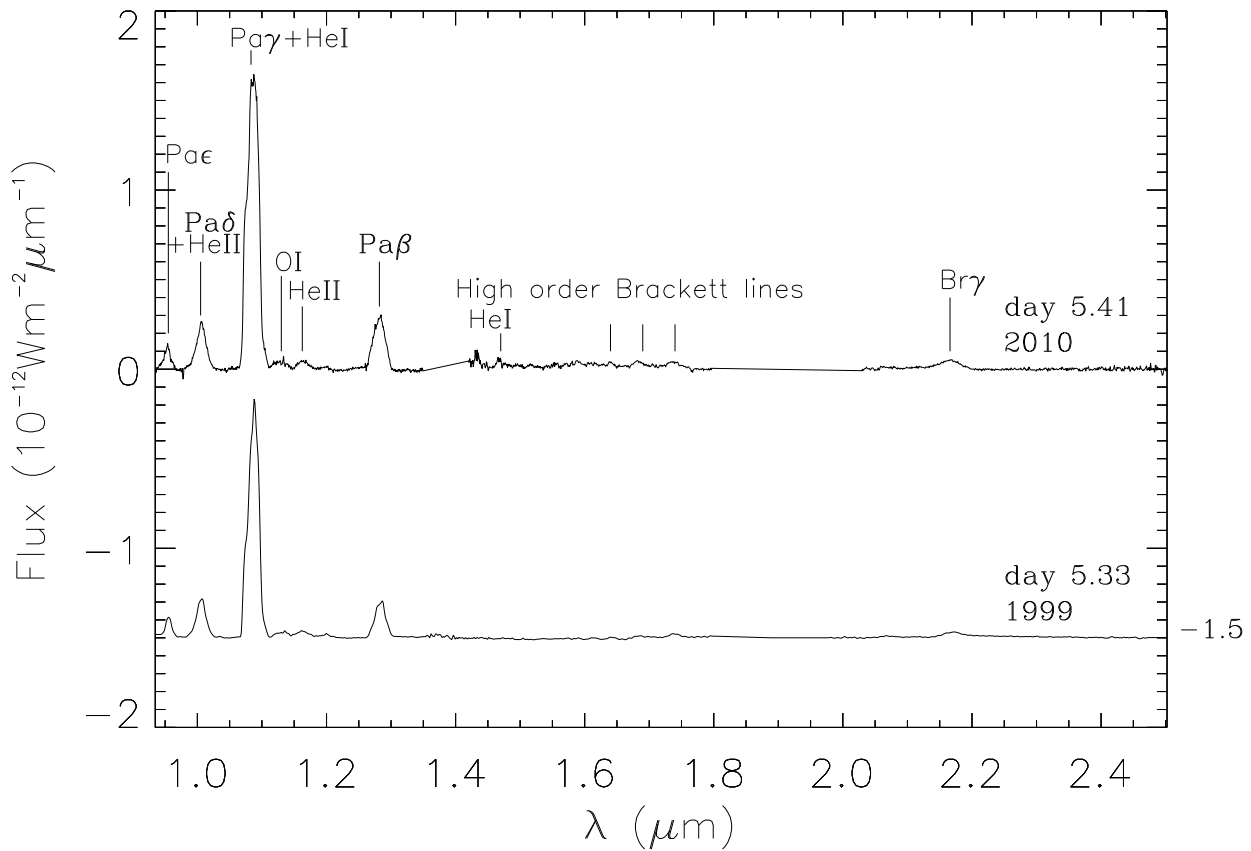}
\caption{Top: IR spectrum of U Sco taken at the NTT on day 5.41 of 2010 outburst. Bottom: IR spectrum of U Sco taken at the NTT 5.33 days after the maximum of the 1999 outburst from \protect\cite{evans}, offset as indicated on right hand side.}
\end{figure*}
\end{center}

\begin{center}
\begin{figure*}
\includegraphics{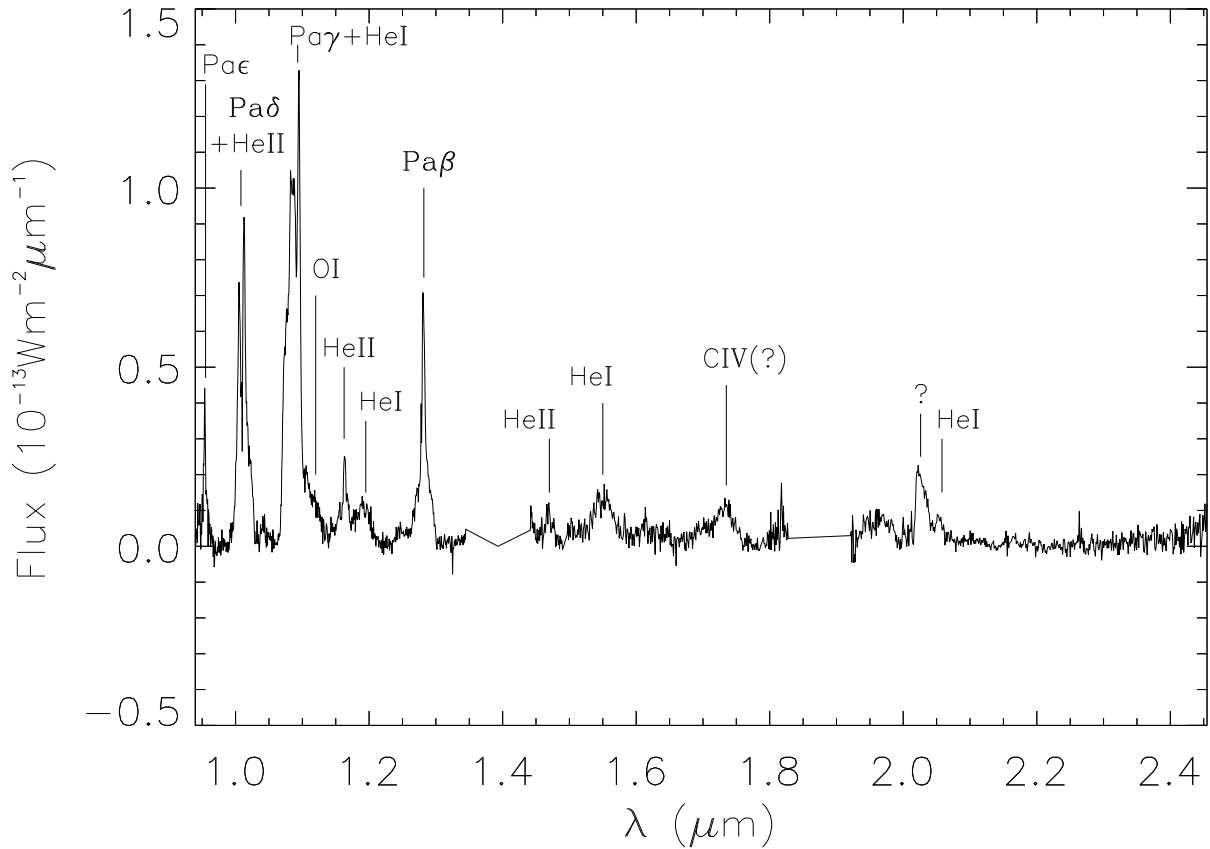}
\caption{IR spectrum of U Sco taken at the NTT on day 9.43 of 2010 outburst.}
\end{figure*}
\end{center}

\begin{center}
\begin{table*}
\begin{minipage}{150mm}

\caption{Broad Component Line Widths (FWZI, kms$^{-1}$) from LT (days ending .81) and SAAO (days ending .93).}
\begin{tabular}{lccccccc} \hline
&    H$\gamma$& N {\sc iii} 4616{\AA} &He {\sc ii} 4686{\AA} &H$\beta$ &H$\alpha$ &He {\sc i} 6678{\AA} &He {\sc i} 7065{\AA} \\ \hline
1.93 & -       	      &	-       	& -	  	     & 8000$\pm$2000 & 10000$\pm$1000	& -		& - \\
4.93 & -	      &	-		& -		     & 9000$\pm$2000 & 9000$\pm$1000	& -		& - \\
5.93 & 12000$\pm$2000 &	-		& -		     & 12000$\pm$2000& 9000$\pm$1000	& -		& - \\
6.81 & 9000$\pm$3000  & 8000$\pm$2000	& 6000$\pm$2000	     & 9000$\pm$2000 & 8000$\pm$1000	& 2000$\pm$500	& 13000$\pm$3000 \\
7.81 & -	      & 8000$\pm$1000	& 6000$\pm$2000	     & 9000$\pm$2000 & 8000$\pm$1000	& 1000$\pm$500	& 10000$\pm$2000 \\
7.93 & 10000$\pm$2000 &	-		& 6000$\pm$3000	     & 9000$\pm$2000 & 9000$\pm$1000 	& -		& - \\
8.81 & 8000$\pm$2000  & 12000$\pm$2000	& 4000$\pm$2000	     & 8000$\pm$2000 & 8000$\pm$500	& 1500$\pm$250	& 12000$\pm$2000 \\
8.93 & 9000$\pm$2000  &	-		& 6000$\pm$2000	     & 9000$\pm$1000 & 9000$\pm$1000	& -		& - \\ 
9.93 & 4000$\pm$1000  &	-		& 4000$\pm$1000	     & 3000$\pm$1000 & 9000$\pm$2000	& -		& - \\
10.93 & 4000$\pm$1000 &	-		& 2000$\pm$500	     & 3000$\pm$1000 & 9000$\pm$1000	& -		& - \\
11.81 & -	      &	-		& 1500$\pm$500	     & 2000$\pm$1000 & 8000$\pm$2000	& -		& - \\
11.93 & 4000$\pm$1000 &	-		& 2000$\pm$500	     & 2000$\pm$500  & 9000$\pm$1000	& -		& - \\
12.81 & -	      &	-		& 1200$\pm$250	     & 1500$\pm$500  & 8000$\pm$1000	& -		& - \\ \hline

\end{tabular}
\end{minipage}
\end{table*}
\end{center}

\subsection{Line Widths and Flux Ratios}
Here we measure the FWZI of each line present in our spectra. Some lines also show narrow components; in these cases we also measure the FWHM of that component. The contribution of instrumental resolution was not found to be significant. Velocities and the associated errors were measured by fitting Gaussian profiles and visual inspection of where the emission meets the continuum. The H$\alpha$ velocity is consistent with previous outbursts at similar times \citep{iijima} at $8000-9000$\,kms$^{-1}$. The width of each line can be seen in Tables 2-4, and can be seen to be changing with time, however the profile of the lines is changing rapidly as can be seen in Figures 1 and 2. Tables 3 and 4 show the evolution of the narrow components of H$\alpha$ and H$\beta$. Such high velocity outflows ($\simeq10,000$\,kms$^{-1}$) were also seen in early ($\leq$5\,days) spectra of the 2010 outburst of U Sco \citep{banerjee}.

Line fluxes were measured by fitting Gaussian profiles to the lines with a least squares fit and can be seen in Table 5. Multiple profiles were used in some cases; an example fit can be seen in Figure 5. These fluxes were then ratioed and compared to theoretical values from \cite{hummer}. From these ratios the abundance of species A to species B can be found using the equation 
\begin{equation}
\frac{F_A}{F_B} = \frac{\lambda_A}{\lambda_B} \times \frac{N_A}{N_B} \times \frac{\alpha_B}{\alpha_A},
\end{equation}
where $F$ is the measured dereddened flux, $N$ is the abundance by number, and $\alpha$ is the recombination coefficient from \cite{hummer} for species A and B respectively. We assume that the sources of H\,{\sc i}, He\,{\sc i}, and He\,{\sc ii} emission are co-extensive. 

This process was repeated for each available He\,{\sc i} (for He$^+$) and He\,{\sc ii} (for He$^{++}$) to hydrogen line ratio. Since our spectra show both ionised states of helium in emission the total abundance can be directly measured instead of using the Saha equation to predict the abundance of one species from the other.

\begin{center}
\begin{table*}
\begin{minipage}{140mm}
\caption{Narrow Component Line Widths (FWHM, kms$^{-1}$) from LT data.}
\begin{tabular}{lccccc}\hline
Day  & 6.81 & 7.81 & 8.81 & 11.81 & 12.81 \\ \hline
H$\beta$ & 1001.95 $\pm$ 72.36 & 836.21 $\pm$ 131.3 & 953.53 $\pm$ 41.62 & 882.85 $\pm$ 61.97 & 1234.63 $\pm$ 85.49 \\
H$\alpha$ & 1071.35 $\pm$ 50.0 & 1146.4 $\pm$ 46.77 & 849.94 $\pm$ 23.94 & 753.96 $\pm$ 57.1 & 1380.0 $\pm$ 41.95 \\ \hline

\end{tabular}
\end{minipage}
\end{table*}
\end{center}

\begin{center}
\begin{table*}
\begin{minipage}{160mm}
\caption{Narrow Component Line Widths (FWHM, kms$^{-1}$) from SAAO data.}
\begin{tabular}{lcccccccc}\hline
Day  & 1.93 & 4.93 & 5.93 & 7.93 & 8.93 & 9.93 & 10.93 & 11.93 \\ \hline
H$\beta$ & -  & - & - & 1782.15$\pm$106.9 & 1318.17$\pm$108.8 & 975.48$\pm$71.76 & 764.72$\pm$53.5 & 1192.30$\pm$80.41\\
H$\alpha$ & - & - & - & 1413.54$\pm$68.99 & 876.78$\pm$40.20 & 778.07$\pm$32.47 & 841.79$\pm$30.53 & 882.26$\pm$36.28\\ \hline

\end{tabular}
\end{minipage}
\end{table*}
\end{center}

\begin{center}
\begin{table*}
\begin{minipage}{160mm}
\caption{Dereddened line fluxes. Fluxes and errors are relative to H$\beta$ on that date for optical data and Pa$\beta$ for near-IR data.}
\begin{tabular}{lcccccccc} \hline
&    H$\beta$&		   H$\gamma$	 &Pa$\beta$	 &Pa$\epsilon$& He {\sc i} 6678{\AA} &He {\sc ii} 4686{\AA} &He {\sc ii} 5411{\AA} & He {\sc ii} 1.163$\mu$m \\ \hline
8.81 & 1$\pm$0.01  & 1.02$\pm$0.03	& -	     & - 	        & 0.007$\pm$0.001	& -	& - & - \\
9.43 & -	   & -			& 1$\pm$0.04 & 0.45$\pm$0.02   	&-			&-&-& 0.18$\pm$0.01 \\	
9.93 & 1$\pm$0.04  & 0.29$\pm$0.02	& -	     & -		 & -			& 0.34$\pm$0.02 & 0.074$\pm$0.010 & - \\
10.93 & 1$\pm$0.02 & 0.12$\pm$0.01	& -	     & -		  & -			& 0.19$\pm$0.01 & 0.022$\pm$0.002 & - \\
11.93 & 1$\pm$0.03 & 0.25$\pm$0.04	& -	     & -		   & -			& 0.28$\pm$0.03 & 0.036$\pm$0.006 & - \\ \hline
\end{tabular}
\end{minipage}
\end{table*}
\end{center}

\subsection{Reddening}

Several studies \citep{barlow,amores,burstein} as well as the dust maps of \cite{schlegel} have shown the reddening of U Sco to be in the range $E(B-V) = 0.09 - 0.36$. The ratios of line fluxes were estimated and compared to the theoretical values derived from \cite{hummer}. We use the optical spectra taken on day 8.81 at LT and day 9.43 at CTIO (Figs 1 and 2), and the IR spectrum taken on day 9.43 at NTT (Fig 4) to estimate the reddening, as by this time the flux ratios are converging on case B values. We use the extinction law and assumption of $R=3.1$ of \cite{howarth}. Using the ratios H$\beta$/Pa$\epsilon$, H$\beta$/Pa$\beta$, H$\gamma$/Pa$\epsilon$, and H$\gamma$/Pa$\beta$ the reddening was found to be in the range $E(B-V) = 0.0 - 0.29$ with a mean of $E(B-V) = 0.14\pm0.12$, consistent with the previous studies. Although U Sco is at a distance of 12\,kpc, low reddening is consistent with both the line of sight leaving the plane of the galaxy and the system being at a height of $z=4.5$\,kpc above the galactic plane \citep{schaeferlong}. For the spectra used in this work we adopt $E(B-V) = 0.2$ as a good compromise between our own value and previous estimates.

\subsection{Helium Abundance}
The helium abundance of U Sco was calculated using H\,{\sc i}, He\,{\sc i}, and He\,{\sc ii} recombination lines. We use the line fluxes on days 8.81 (LT), 9.93-11.93 (SAAO), and 9.43 (NTT) to estimate the helium abundance, as by this time the line ratios are converging on case B values. The large H$\alpha$/H$\beta$ ratio throughout the time coverage we have available shows that H$\alpha$ should not be used in our abundance analysis due to optical depth effects.

Abundance analyses of helium are complicated by the metastability of the lowest triplet level of He\,{\sc i}, 2$^3$S, which can cause some lines to become optically thick. Furthermore, collisional excitations from this level enhance triplet lines. The collisional contributions are 56\% for 5876{\AA}, 78\% for 7065{\AA}, and 72\% for 1.083$\mu$m using relations from \cite{kingdon} and \cite{peimbert} at a temperature of $2 \times 10^{4}$\,K at the high density limit. The contribution is lower in the singlet lines, although collisions from the 2$^1$S level must be taken into account. The ratio of 7065/4471 is an optical depth indicator \citep{osterbrock} and this ratio (1.31) is inconsistent with the case B value ($\sim0.5$), further evidence that the 7065{\AA} line is unsuitable for this analysis. In our abundance analysis we consider the singlet He\,{\sc I} line at 6678{\AA} for which the collisional contribution is 27\% using equation 12 of \cite{kingdon} at a temperature of $2 \times 10^{4}$\,K. At temperatures of $10^{4}$ and $3 \times 10^{4}$ the collisional contributions for this line are 7\% and 36\% respectively. We also use He\,{\sc II} lines at 4687{\AA} (LT and SAAO), 5411{\AA} (SAAO), and 1.163$\mu$m (NTT).

Following \cite{evans} we estimate the electron density using Paschen and Brackett series H\,{\sc i} lines from our near-IR spectra (Figs 4 and 5) using the equation:
\begin{equation}
F(H) = \frac{\epsilon \phi N_{e} ^{2}}{\alpha} \times \frac{V^{3} t^{3}}{3D^{2}},
\end{equation}
where $F$ is the flux of a hydrogen line, $\epsilon$ is the case B emissivity from \cite{hummer}, $\alpha$ is the number of electrons per hydrogen, $\phi$ is the volume filling factor, $N_e$ is the electron density, $V$ is the velocity, $D$ is the distance, and $t$ is the time since maximum in days. We follow \cite{evans} by assuming $\alpha=1$, $\phi=0.01$, and $D=12$\,kpc. Our results are consistent with those of \cite{evans} at similar times, therefore we adopt their values of electron density of 10$^{10}$\,cm$^{-3}$ for data taken before day 9.43 and 10$^{9}$\,cm$^{-3}$ for data taken at later times. The recombination coefficients are relatively insensitive to temperature; we assume $T_e = 20,000$\,K.
We derive $N$(He$^+$)/$N$(H$^+$) = $0.012\pm0.015$ and $N$(He$^{++}$)/$N$(H$^+$) = $0.061\pm0.010$ from the data obtained at LT, NTT, and CTIO using the He\,{\sc i} line at 6678{\AA} and the He\,{\sc ii} line at 1.163$\mu$m. From the SAAO data we derive $N$(He$^{++}$)/$N$(H$^+$) = $0.047\pm0.011$ using the He\,{\sc ii} line at 4686{\AA} and $N$(He$^{++}$)/$N$(H$^+$) = $0.076\pm0.023$ using the line at 5411{\AA}. These results are summarised in Table 6.

\begin{center}
\begin{table}
\caption{Helium abundances}
\begin{tabular}{lc}\hline
Helium Line & Derived Abundance\\ \hline
He\,{\sc i} 6678{\AA} & $0.012 \pm 0.015$ \\
He\,{\sc ii} 4686{\AA} & $0.047 \pm 0.011$ \\
He\,{\sc ii} 5411{\AA} & $0.076 \pm 0.023$ \\
He\,{\sc ii} 1.163$\mu$m & $0.061 \pm 0.010$ \\ \hline
\end{tabular}
\end{table}
\end{center}

\begin{center}
\begin{figure}
\includegraphics[scale=0.5]{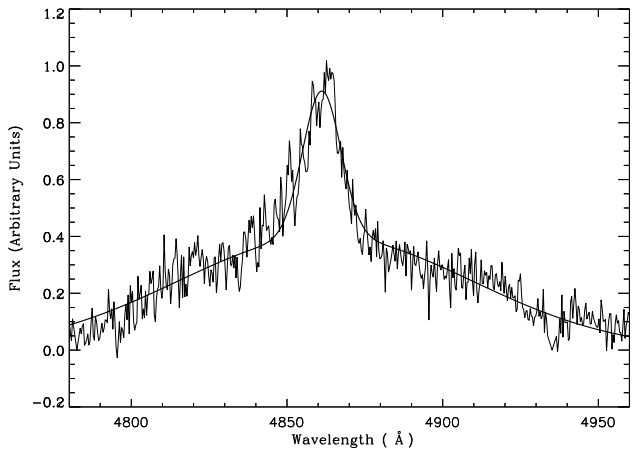}
\caption{Fit of H$\beta$ using a pair of Gaussian profiles.}
\end{figure}
\end{center}

\section{Discussion}
The helium abundance in U Sco has been the subject of many studies which have resulted in a wide range of values from 0.16 to 4.5 \citep{iijima,barlow,anupama,evans}. We find the abundance to be $N$(He)/$N$(H)$ = 0.073\pm0.031$, lower than that found by \cite{iijima}, \cite{barlow}, \cite{anupama}, and \cite{evans}. A high helium abundance would suggest that the secondary in U Sco is a highly evolved star, however the value derived here is close to the solar value N(He)/N(H) = 0.085 \citep{asplund}. Our estimate suggests that assumptions about the helium-rich nature of the secondary are unfounded and that the secondary did not accrete helium-rich material significantly in the post common envelope phase \citep{hachisukato}. In the evolutionary calculations of \cite{sarna}, U Sco is best represented by their sequence B, with a hydrogen-rich secondary. For our helium abundance analysis we have considered the data ($\ge$day 8) and the lines for which we can be confident that case B recombination applies. Other estimates have been larger than our value, especially those based only on the He\,{\sc ii} lines. These large abundances result from the use of the Saha equation, which is highly sensitive to the assumed electron temperature. However, we use both He\,{\sc i} and He\,{\sc ii} lines in our analysis, avoiding the use of the equation to derive the total helium abundance. The temperature dependencies in this analysis are in the recombination coefficients, which are relatively insensitive to the value chosen, and in the collisional correction for He\,{\sc i} lines. Using a range of lines from days 8.81 and 9.43 and a range of recombination coefficients we determine several He$^{++}$/He$^{+}$ ratios, each of which suggest electron temperatures of $2.3-2.4 \times 10^4$\,K, therefore we use recombination coefficients for the closest temperature from \cite{hummer} of $T_e =2 \times 10^4$\,K.

The hydrogen line profiles have broad and narrow components. The narrow components become prominent around day 8 and arise from slow moving material at speeds of $\sim1000$\,kms$^{-1}$. The broad components are composed of material travelling at velocities of up to $\sim10,000$\,kms$^{-1}$. The rise in the strength of the narrow components could be explained by optical depth effects. The ejecta begins to become optically thin around day 8; before this time slower moving material would be concealed by the shell of optically thick fast moving material. As the faster moving material expands and becomes optically thin the flux from the slow moving material becomes visible. He\,{\sc ii} lines also begin to dominate the helium emission around day 8, however this is most likely due to photo-ionisation of helium rather than an optical depth effect. The measured line widths are generally higher for the SAAO data than for near-simultaneous LT data; possible reasons for this are the difference in resolution and signal to noise issues. There also appears to be clear structure to the H$\beta$ line in the LT spectra which is not present in the SAAO spectra. A possible source of the narrow components is an inclination effect; since U Sco is a high inclination system most of the velocity is in the plane of the sky, therefore we may be observing a small radial component. This is consistent with the asymmetric ejecta models of \cite{drake}. Data over a longer time range and with better orbital phase coverage is required to explore this further.

\section{Conclusion}
We present optical and near-infrared observations of the 2010 outburst of U Scorpii. We find the helium abundance of U Scorpii to be $N$(He)/$N$(H)$ = 0.073\pm0.031$. This estimate is lower than most previous studies and does not support their conclusions which suggest that the secondary in this system is helium-rich. The velocities seen in the 2010 outburst are consistent with those seen in previous outbursts, with some hydrogen and He\,{\sc ii} lines being seen to have both narrow and broad components in our later spectra. Further observations are planned to investigate the nature and metallicity of the secondary in this system. 

\section{Acknowledgements}
The Liverpool Telescope is operated on the island of La Palma by Liverpool John Moores University in the Spanish Observatorio del Roque de los Muchachos of the Instituto de Astrofisica de Canarias with financial support from the UK Science and Technology Facilities Council. M.B.N.K. was supported by the Peter and Patricia Gruber Foundation through the IAU-PPGF fellowship, by the Peking University One Hundred Talent Fund (985), and by the National Natural Science Foundation of China (grants 11010237 and 11043007). This paper uses observations made at the South African Astronomical Observatory (SAAO). The authors would like to thank the anonymous referee for their helpful comments.

\bibliographystyle{mn2e}
\bibliography{refs}

\begin{thebibliography}{}

\bibitem[\protect\citeauthoryear{{Am{\^o}res} \& {L{\'e}pine}}{{Am{\^o}res} \&
  {L{\'e}pine}}{2005}]{amores}
{Am{\^o}res} E.~B.,  {L{\'e}pine} J.~R.~D.,  2005, AJ, 130, 659

\bibitem[\protect\citeauthoryear{{Anupama} \& {Dewangan}}{{Anupama} \&
  {Dewangan}}{2000}]{anupama}
{Anupama} G.~C.,  {Dewangan} G.~C.,  2000, AJ, 119, 1359

\bibitem[\protect\citeauthoryear{{Asplund}, {Grevesse}, {Sauval} \&
  {Scott}}{{Asplund} et~al.}{2009}]{asplund}
{Asplund} M.,  {Grevesse} N.,  {Sauval} A.~J.,    {Scott} P.,  2009, ARAA, 47,
  481

\bibitem[\protect\citeauthoryear{{Banerjee}, {Das}, {Ashok}, {Rushton},
  {Eyres}, {Maxwell}, {Worters}, {Evans} \& {Schaefer}}{{Banerjee}
  et~al.}{2010}]{banerjee}
{Banerjee} D.~P.~K.,  {Das} R.~K.,  {Ashok} N.~M.,  {Rushton} M.~T.,  {Eyres}
  S.~P.~S.,  {Maxwell} M.~P.,  {Worters} H.~L.,  {Evans} A.,    {Schaefer}
  B.~E.,  2010, MNRAS, 408, L71

\bibitem[\protect\citeauthoryear{{Barlow}, {Brodie}, {Brunt}, {Hanes}, {Hill},
  {Mayo}, {Pringle}, {Ward}, {Watson}, {Whelan} \& {Willis}}{{Barlow}
  et~al.}{1981}]{barlow}
{Barlow} M.~J.,  {Brodie} J.~P.,  {Brunt} C.~C.,  {Hanes} D.~A.,  {Hill} P.~W.,
   {Mayo} S.~K.,  {Pringle} J.~E.,  {Ward} M.~J.,  {Watson} M.~G.,  {Whelan}
  J.~A.~J.,    {Willis} A.~J.,  1981, MNRAS, 195, 61

\bibitem[\protect\citeauthoryear{{Burstein} \& {Heiles}}{{Burstein} \&
  {Heiles}}{1982}]{burstein}
{Burstein} D.,  {Heiles} C.,  1982, AJ, 87, 1165

\bibitem[\protect\citeauthoryear{{Diaz}, {Williams}, {Luna}, {Moraes} \&
  {Takeda}}{{Diaz} et~al.}{2010}]{diaz}
{Diaz} M.~P.,  {Williams} R.~E.,  {Luna} G.~J.,  {Moraes} M.,    {Takeda} L.,
  2010, ArXiv e-prints

\bibitem[\protect\citeauthoryear{{Drake} \& {Orlando}}{{Drake} \&
  {Orlando}}{2010}]{drake}
{Drake} J.~J.,  {Orlando} S.,  2010, APJL, 720, L195

\bibitem[\protect\citeauthoryear{{Evans}, {Krautter}, {Vanzi} \&
  {Starrfield}}{{Evans} et~al.}{2001}]{evans}
{Evans} A.,  {Krautter} J.,  {Vanzi} L.,    {Starrfield} S.,  2001, AAP, 378,
  132

\bibitem[\protect\citeauthoryear{{Hachisu}, {Kato}, {Nomoto} \&
  {Umeda}}{{Hachisu} et~al.}{1999}]{hachisukato}
{Hachisu} I.,  {Kato} M.,  {Nomoto} K.,    {Umeda} H.,  1999, ApJ, 519, 314

\bibitem[\protect\citeauthoryear{{Hanes}}{{Hanes}}{1985}]{hanes}
{Hanes} D.~A.,  1985, MNRAS, 213, 443

\bibitem[\protect\citeauthoryear{{Howarth}}{{Howarth}}{1983}]{howarth}
{Howarth} I.~D.,  1983, MNRAS, 203, 301

\bibitem[\protect\citeauthoryear{{Hummer} \& {Storey}}{{Hummer} \&
  {Storey}}{1987}]{hummer}
{Hummer} D.~G.,  {Storey} P.~J.,  1987, MNRAS, 224, 801

\bibitem[\protect\citeauthoryear{{Iijima}}{{Iijima}}{2002}]{iijima}
{Iijima} T.,  2002, AAP, 387, 1013

\bibitem[\protect\citeauthoryear{{Kingdon} \& {Ferland}}{{Kingdon} \&
  {Ferland}}{1995}]{kingdon}
{Kingdon} J.,  {Ferland} G.~J.,  1995, APJ, 442, 714

\bibitem[\protect\citeauthoryear{{Morales-Rueda}, {Carter}, {Steele}, {Charles}
  \& {Worswick}}{{Morales-Rueda} et~al.}{2004}]{morales}
{Morales-Rueda} L.,  {Carter} D.,  {Steele} I.~A.,  {Charles} P.~A.,
  {Worswick} S.,  2004, Astronomische Nachrichten, 325, 215

\bibitem[\protect\citeauthoryear{{Osterbrock} \& {Ferland}}{{Osterbrock} \&
  {Ferland}}{2006}]{osterbrock}
{Osterbrock} D.~E.,  {Ferland} G.~J.,  2006, Astrophysics of Gaseous Nebulae
  and Active Galactic Nuclei.
University Science Books

\bibitem[\protect\citeauthoryear{{Pagnotta}, {Schaefer}, {Landolt}, {Clem},
  {Schlegel}, {Page}, {Osborne}, {Handler} \& {Walter}}{{Pagnotta}
  et~al.}{2011}]{pagnotta}
{Pagnotta} A.~S.,  {Schaefer} B.~E.,  {Landolt} A.~U.,  {Clem} J.~L.,
  {Schlegel} E.~M.,  {Page} K.~L.,  {Osborne} J.~P.,  {Handler} G.,    {Walter}
  F.~M.,  2011, in American Astronomical Society Meeting Abstracts \#217
  Vol.~43 of Bulletin of the American Astronomical Society, {The 2010 Eruption
  of the Recurrent Nova U Scorpii}.
pp 338.14--+

\bibitem[\protect\citeauthoryear{{Peimbert} \& {Torres-Peimbert}}{{Peimbert} \&
  {Torres-Peimbert}}{1987}]{peimbert}
{Peimbert} M.,  {Torres-Peimbert} S.,  1987, RMXAA, 15, 117

\bibitem[\protect\citeauthoryear{{Sarna}, {Ergma} \& {Gerskevits}}{{Sarna}
  et~al.}{2006}]{sarna}
{Sarna} M.~J.,  {Ergma} E.,    {Gerskevits} J.,  2006, ACTAA, 56, 65

\bibitem[\protect\citeauthoryear{{Schaefer}}{{Schaefer}}{2004}]{schaefer04}
{Schaefer} B.~E.,  2004, IAU Circ, 8279, 3

\bibitem[\protect\citeauthoryear{{Schaefer}}{{Schaefer}}{2010}]{schaeferlong}
{Schaefer} B.~E.,  2010, APJS, 187, 275

\bibitem[\protect\citeauthoryear{{Schaefer}, {Harris}, {Dvorak}, {Templeton} \&
  {Linnolt}}{{Schaefer} et~al.}{2010}]{schaefer}
{Schaefer} B.~E.,  {Harris} B.~G.,  {Dvorak} S.,  {Templeton} M.,    {Linnolt}
  M.,  2010, IAU Circ, 9111, 1

\bibitem[\protect\citeauthoryear{{Schaefer} \& {Ringwald}}{{Schaefer} \&
  {Ringwald}}{1995}]{schaefer&ringwald}
{Schaefer} B.~E.,  {Ringwald} F.~A.,  1995, APJL, 447, L45+

\bibitem[\protect\citeauthoryear{{Schlegel}, {Finkbeiner} \&
  {Davis}}{{Schlegel} et~al.}{1998}]{schlegel}
{Schlegel} D.~J.,  {Finkbeiner} D.~P.,    {Davis} M.,  1998, APJ, 500, 525

\bibitem[\protect\citeauthoryear{{Starrfield}}{{Starrfield}}{2008}]{starrfield}
{Starrfield} S.,  2008, Classical Novae.
Cambridge University Press

\bibitem[\protect\citeauthoryear{{Starrfield}, {Sonneborn}, {Sparks}, {Shaviv},
  {Williams}, {Heathcote}, {Ferland}, {Gehrz}, {Ney} \& {Kenyon}}{{Starrfield}
  et~al.}{1988}]{Starrfield2}
{Starrfield} S.,  {Sonneborn} G.,  {Sparks} W.~M.,  {Shaviv} G.,  {Williams}
  R.~E.,  {Heathcote} S.,  {Ferland} G.,  {Gehrz} R.~D.,  {Ney} E.~P.,
  {Kenyon} S.,  1988, in ESA Special Publication Vol.~281 of ESA Special
  Publication, {Observations and simulations of recurrent novae: U Sco and V394
  CrA}.
pp 167--170

\bibitem[\protect\citeauthoryear{{Steele}, {Smith}, {Rees}, {Baker}, {Bates},
  {Bode}, {Bowman}, {Carter}, {Etherton}, {Ford}, {Fraser} \&
  {Gomboc}}{{Steele} et~al.}{2004}]{steele}
{Steele} I.~A.,  {Smith} R.~J.,  {Rees} P.~C.,  {Baker} I.~P.,  {Bates} S.~D.,
  {Bode} M.~F.,  {Bowman} M.~K.,  {Carter} D.,  {Etherton} J.,  {Ford} M.~J.,
  {Fraser} S.~N.,    {Gomboc} A.,  2004, in {J.~M.~Oschmann Jr.} ed., Society
  of Photo-Optical Instrumentation Engineers (SPIE) Conference Series Vol.~5489
  of Presented at the Society of Photo-Optical Instrumentation Engineers (SPIE)
  Conference, {The Liverpool Telescope: performance and first results}.
pp 679--692

\bibitem[\protect\citeauthoryear{{Thoroughgood}, {Dhillon}, {Littlefair},
  {Marsh} \& {Smith}}{{Thoroughgood} et~al.}{2001}]{thoroughgood}
{Thoroughgood} T.~D.,  {Dhillon} V.~S.,  {Littlefair} S.~P.,  {Marsh} T.~R.,
  {Smith} D.~A.,  2001, MNRAS, 327, 1323

\bibitem[\protect\citeauthoryear{{Truran}, {Livio}, {Hayes}, {Starrfield} \&
  {Sparks}}{{Truran} et~al.}{1988}]{truran}
{Truran} J.~W.,  {Livio} M.,  {Hayes} J.,  {Starrfield} S.,    {Sparks} W.~M.,
  1988, ApJ, 324, 345

\bibitem[\protect\citeauthoryear{{Worters}, {Eyres}, {Rushton} \&
  {Schaefer}}{{Worters} et~al.}{2010}]{worters}
{Worters} H.~L.,  {Eyres} S.~P.~S.,  {Rushton} M.~T.,    {Schaefer} B.,  2010,
  IAU Circ, 9114, 1

\bibitem[\protect\citeauthoryear{{Yaron}, {Prialnik}, {Shara} \&
  {Kovetz}}{{Yaron} et~al.}{2005}]{yaron}
{Yaron} O.,  {Prialnik} D.,  {Shara} M.~M.,    {Kovetz} A.,  2005, APJ, 623,
  398

\end{thebibliography}

\end{document}